\documentclass{eptcs}
\providecommand{\event}{ACL2 2017} 
\usepackage{underscore}           
\usepackage{fancyvrb,graphicx,latexsym,pifont,amsmath,amstext,amssymb,amsfonts,epic,eepic}

\newtheorem{theorem*}{Theorem}
\newtheorem{lemma}{Lemma}[section]
\newtheorem{definition}{Definition}[section]

\newtheorem{computation}{Computation}

\numberwithin{equation}{section}

\title{A Computationally Surveyable Proof\\ of the Group Properties of an Elliptic Curve}
\author{David M. Russinoff
\email{david@russinoff.com}}
\def\titlerunning{Group Properties of an Elliptic Curve}
\def\authorrunning{D.M. Russinoff}

\begin{document}
\maketitle

\begin{abstract}
We present an elementary proof of the abelian group properties of the elliptic curve
known as {\it Curve25519}, as a component of a comprehensive proof of correctness of a hardware implementation of the
associated Diffie-Hellman key agreement algorithm.  The entire proof has
been formalized and mechanically verified with ACL2, and is {\it computationally surveyable}
in the sense that all steps that require mechanical support are presented in such a way that they
may be readily reproduced in any suitable programming language.
\end{abstract}

\section{Introduction}

An effort is under way at Intel to develop and verify a formal model and a hardware implementation of the elliptic curve
key agreement algorithm known as {\it Curve25519} \cite{bernstein}, using ACL2.  The most
challenging aspect of this problem is the proof of the abelian group properties of the curve, especially 
associativity.  This result may be viewed either as a deep theorem of algebraic or projective
geometry \cite{hank,bernstein2}, accessible only to experts in that field, or as an elementary but computationally intensive
arithmetic exercise, involving, as Bernstein \cite{bernstein} observes, ``standard (but lengthy) calculations''.  Silverman and 
Tate \cite{tate} attempt to quantify the effort with a ``tongue-in-cheek estimate'':

\begin{quote}
Of course, there are a lot of cases to consider \dots.  But in a few days you will be able to check associativity
using these formulas.  So we need say nothing more about the proof of the associative law!
\end{quote}
What remains to be said is that there is compelling evidence (see below) that an elementary hand proof of this result is a practical 
impossibility.  The first serious attack on the problem, by Friedl \cite{friedl}, was a combination of mathematical analysis 
and symbolic computation with the CoCoA (Computations in Commutative Algebra) package.  Building on Friedl's results,
Th\'ery \cite{thery} later developed a comprehensive formal proof with Coq.  (Both papers address the somewhat more general class of 
Weierstrass curves rather than the one on which we focus here, but there is no difference in computational complexity.)
These two efforts, which together represent a significant achievement, may be contrasted in the terminology of automated 
reasoning \cite{autarky}: Friedl's work is {\it accepting} 
insofar as it treats CoCoA as a trusted oracle, whereas Th\'ery's proof is {\it autarkic} by virtue of performing all logical deductions 
and supporting computations within the same formal system.

From a traditional mathematical perspective, however, both of these results are open to the same common criticism of
computer-assisted proofs.  There is general agreement in the mathematical community that it is desirable for a proof to be 
{\it surveyable} in the sense that each of its assertions could be derived manually by a competent reader as a logical consequence of 
preceding assertions and otherwise established results, and that the proof is short enough to be comprehended.  One goal behind this 
principle is correctness, but equally important is the desire for mathematical growth---the propagation and advancement of techniques 
and ideas.

In the realm of computing, this is often an unattainable objective---reliance on a mechanical proof assistant may be unavoidable.  It is common to find in a published proof in this field, in lieu of a cogent argument, an appeal to the authority of an established proof system.  This device is a stark realization of Tymoczko's allegory of the infallible Martian genius whose proclamations go unquestioned---proof by ``Simon Says'' \cite{tymo}.  It may provide evidence of correctness but does little to illuminate the underlying mathematics.

Dependence on mechanical assistance, however, need not preclude a full exposition of a proof.  For example, the correctness of a hardware divider typically depends on a relation between the value and indices of each entry of an array that is too large to be either generated or checked by hand, but it should be possible to characterize the computation in such a way that it can be understood and machine-checked by the skeptical reader.

This suggests a judicious weakening of the conventional notion of surveyability.  A proof may be said to be {\it computationally surveyable} if its only departure from that notion is its dependence on unproved assertions that satisfy the following criteria:
\begin{itemize}

\item[(1)] The assertion pertains to a function for which a clear constructive definition has been provided, and merely specifies the value of that function corresponding to a concrete set of arguments.

\item[(2)] The computation of this value has been performed mechanically by the author of the proof in a reasonably short time.

\item[(3)] A competent reader could readily code the function in the programming language of his choice and verify the asserted result on his own computing platform.

\end{itemize}
Such a proof, though still objectionable to those who insist on strict surveyability, can convey a comprehensive understanding of a theorem and is susceptible to a process of social review, thus oppugning a commonly stated basis for the objection.

Neither of the treatments of the elliptic curve group properties cited above attempts such a proof, perhaps because the supporting tool or its application to the problem at hand is too complicated to admit a concise specification.  Thus, Friedl simply attributes unproved results to CoCoA, while Th\'ery's claims depend on an undisclosed ``tactic'' that has reportedly been implemented in Coq.

An integrated computationally surveyable proof of a result of this sort, which combines subtle mathematical analysis with intensive computation, is best carried out with the support of an interactive prover based on an efficient executable logic.
We shall present such a proof of this theorem that has been formalized and mechanically checked in its entirety with ACL2.
The computational results for which we rely solely on ACL2 for verification, as opposed to proof checking, (all of which are in Section~\ref{axioms}) are labeled as {\bf Computation}s, and are thus clearly distinguished from other steps in the proof, which are listed as {\bf Lemma}s.  All computations are performed on S-expressions and are most naturally performed in LISP, but can be readily implemented in any language that provides linked lists.  Moreover, our exposition is confined to conventional mathematical terminology and notation, with no reference to the ACL2 logic.

Our proof benefits significantly from the two earlier efforts, both in its overall approach and through its appropriation of specific lemmas.  In particular, we follow \cite{thery} in the representation of polynomials in sparse Horner normal form, using a normalization procedure adapted from \cite{gregoire}.  Furthermore, our Lemmas~\ref{l1},~\ref{l2}, and~\ref{c6} are variants of results found in \cite{thery} (two of which are inherited from \cite{friedl}).

The supporting materials for this paper include several subdirectories of {\tt books/\-proj\-ects} in the ACL2 repository.
The main script resides in {\tt curve25519}.  The basis of Curve25519 is the primality of $\wp = 2^{255}-19$, which is
proved in {\tt quadratic-reci\-procity} by Pratt's method~\cite{pratt}and explained in \cite{prime}.
Fermat's Theorem ($a^{\wp-1} \bmod \wp = 1$ when $a$ is not divisible by $\wp$), which allows the inversion operator in the
field $\mathbb{F}_\wp$ to be defined as $a^{-1} = a^{\wp-2} \bmod \wp$, is also formalized in {\tt quadratic-reciprocity}.

Our formalization of sparse Horner normal forms is in the subdirectory {\tt shnf}.
Following~\cite{gregoire}, we define an efficiently computable normalization of polynomial terms and an
evaluation function on normal forms, and prove equality between the value of a polynomial and that of its representation,
for all variable assignments. Thus, the equivalence of two polynomials we may be established by computing their normalizations
and observing that they coincide.

Of course, the utility of this method rests on the property of completeness: equivalent polynomials always produce
the same representation.  According to the authors of the Coq proof, which does not address this property, it cannot even
be stated within their formal framework.  Our development includes a constructive
proof of this result that we have formalized in ACL2, based on a function that computes, for a given pair of two polynomials,
a list of variable assignments for which the values of the polynomials differ, whenever such a list exists.
This result is not required for our present purpose, but is documented elsewhere~\cite{shnf}.

The distinguishing features of our proof that enable the objective of computational surveyability are (1)~a specialized rewriting procedure that reduces the normal form of a polynomial according to the curve equation (Definition~\ref{reduce}), and (2)~an encoding of group elements as integer triples, which facilitates symbolic computation of the group operation (Definition~\ref{bullet}).  Both of these functions require automated computation but admit concise specifications and correctness proofs.  Furthermore, a modest improvement in efficiency over the more general Coq proof tactic is suggested by a comparison of execution times of the three computational results of \cite{thery} (9.2, 3.9, and 18.8 seconds for {\tt spec3\_assoc}, {\tt spec2\_assoc}, and {\tt spec3\_assoc}, respectively) and our versions of the same computations (3.78, 0.36, and 3.8 seconds for Computations~\ref{p9}, \ref{p10}, and~\ref{p11}). We exploit this facility by performing several more intensive computations, thereby eliminating much of Th\'ery's analysis, which he characterizes as ``really tedious''.  In particular, Computation~\ref{p12}, which is proved in 26.2 seconds, disposes of a critical case of associativity.  It is also worth noting that if the polynomial involved in this result were expanded into a sum of monomials, as might be done in a direct hand proof based on ``standard computations'', the number of terms would exceed $10^{25}$.  Clearly, the reader who completes such a proof ``in a few days'' is exceptionally good with figures.

\section{Curve25519}\label{curve}

Let $\wp = 2^{255}-19$ and $A = 486662$.  The primality of $\wp$ is proved in \cite{prime}.
The field of order $\wp$ is the set $\mathbb{F}_\wp = \{0,1,2,\ldots,\wp-1\}$
with the operations of addition and multiplication modulo $\wp$.
Every $n \in \mathbb{Z}$ naturally corresponds to the field element $n \bmod \wp$, which we denote as $\bar{n}$.
The field operations will be denoted by the usual symbols: if $x \in \mathbb{F}_\wp$, $y \in \mathbb{F}_\wp$, and $k \in \mathbb{N}$,
then ``$x+y$'', ``$x-y$'', ``$-x$'', ``$xy$'', or ``$x^k$'' may refer to an operation in either $\mathbb{F}_\wp$ or $\mathbb{Z}$, 
depending on context, whereas ``$x/y$'' will only denote an operation in $\mathbb{F}_\wp$.

\begin{definition}\label{ec}
$EC = \{(x,y) \in \mathbb{F}_\wp \times \mathbb{F}_\wp \;|\; y^2 = x^3 + Ax^2 + x\} \cup \{\infty\}$.
\end{definition}
Our goal is to show that $EC$ is an abelian group under the following operation:

\begin{definition}\label{grp}
Let $P \in EC$ and $Q \in EC$.
\begin{itemize}

\item[(1)] $P \oplus \infty = \infty \oplus P = P$.

\item[(2)] If $P = (x, y)$, then $P \oplus (x, -y) = \infty$.

\item[(3)] If $P = (x_1, y_1)$, $Q = (x_2, y_2) \neq (x_1, -y_1)$, and
$
\lambda = \left\{\begin{array}{ll}
    \frac{y_2-y_1}{x_2-x_1} & \mbox{if $x_1 \neq x_2$}\\
    \frac{3x_1^2 + 2Ax_1 + 1}{2y_1} & \mbox{if $x_1 = x_2$,}\end{array}\right.
$
then $P \oplus Q = (x,y)$, where
$
x = \lambda^2 - A - x_1 - x_2
$
and
$
y = \lambda(x_1 - x) - y_1.
$

\end{itemize}
\end{definition}
Clearly, $\infty$ is the identity element, the inverse of $P = (x,y)$ is $\ominus P = (x, -y)$, and according to Corollary~1 
of \cite{prime}, the origin $O = (0,0)$ is the only element of order 2.\medskip\\
{\bf Remark}. If we consider Definition~\ref{ec} as an equation over $\mathbb{R}$ instead of $\mathbb{F}_\wp$, then Definition~\ref{grp}
admits a simple geometric interpretation.  Except when $Q = \ominus P$, the line connecting points $P$ and $Q$ on the 
curve (or the tangent line at $P$, in case $P=Q$) intersects the curve at another point, $R$.  If we were to define the operation as 
$P \oplus Q = \ominus R$, then
analytic geometry would yield the formula in the definition.  If $Q = \ominus P$, the third point of intersection is taken to be
$\infty$.\medskip\\
In the sequel, we shall assume that
$P_0 = (x_0, y_0)$, $P_1 = (x_1, y_1)$, and $P_2 = (x_2, y_2)$ are fixed elements of $EC$
that are distinct from $\infty$ (but not necessarily from one another), in order to obviate repetition of such hypotheses.
Any result pertaining to these points may be generalized by replacing them with arbitrary finite points of $EC$.
We begin with two simple consequences of Definition~\ref{grp}.

\begin{lemma}\label{l1}
$P_0 \oplus P_1 \neq P_0$.
\end{lemma}

{\sc Proof}: Suppose $P_0 \oplus P_1 = P_0$.  Equating $y$-coordinates, we have
$y_0 = \lambda(x_0 - x_0) - y_0 = -y_0$,
which implies $2y_0 = 0$ and hence (since $\mathbb{F}_\wp$ is of odd characteristic $\wp$) $y_0 = 0$,  
which implies $x_0 = 0$.  But $x_1$ cannot be 0, as this would imply $P_0 = O \oplus O = \infty$.
Thus, the equation $x_0 = \lambda^2 - A - x_0 - x_1$
reduces to
\[
\frac{y_1^2}{x_1^2} = \lambda^2 = x_1 + A,
\]
which implies $x_1^3 + Ax_1^2 + x_1 = y_1^2 = x_1^3 + Ax_1^2$, contradicting $x_1 \neq 0$.~$\Box$

\begin{lemma}\label{l2}
If $P_0 \oplus P_1 = P_0 \oplus (\ominus P_1)$, then either $P_0 = O$ or $P_1 = O$.
\end{lemma}

{\sc Proof}: If $P_0 = \ominus P_1$, then 
\[
P_0 \oplus P_0 = P_0 \oplus (\ominus P_1) = P_0 \oplus P_1 = \ominus P_1 \oplus P_1 = \infty,
\]
which implies $P_0 = O$.  Similarly, if $P_0 = P_1$, then
\[
P_0 \oplus P_0 = P_0 \oplus P_1 = P_0 \oplus (\ominus P_1) = P_0 \oplus (\ominus P_0) = \infty.
\]
Therefore, we may assume $x_0 \neq x_1$.  Equating the $x$-coordinates of $P_0 \oplus P_1$ and $P_0 \oplus (\ominus P_1)$, we have
\[
\left(\frac{y_1-y_0}{x_1-x_0}\right)^2 - x_0 - x_1 = \left(\frac{y_1+y_0}{x_1-x_0}\right)^2 - x_0 - x_1,
\]
which implies $4y_0y_1 = 0$, and hence either $y_0 = 0$ or $y_1 = 0$.~$\Box$

\section{Encoding Points on the Curve as Integer Triples}\label{triples}

Our scheme for symbolic computation of the group operation is based on a mapping from $\mathbb{Z}^3$ to $\mathbb{F}_\wp^2$:

\begin{definition}
If ${\cal P} = (m,n,z) \in \mathbb{Z}^3$, where $z$ is not divisible by $\wp$, then
\[
\mathit{decode}({\cal P}) = \left(\frac{\bar{m}}{\bar{z}^2}, \frac{\bar{n}}{\bar{z}^3}\right) \in \mathbb{F}_\wp^2
\]
and ${\cal P}$ is said to be an encoding of $\mathit{decode}({\cal P})$.
\end{definition}
Note that every $P  = (x, y) \in \mathbb{F}_\wp^2$ admits the canonical encoding ${\cal P} = (x, y, 1)$.

The motivation for this definition is that an encoding of $P \oplus Q$ can often be readily derived from encodings of $P$ and
$Q$ in certain cases of interest.  We define a partial addition operation on $\mathbb{Z}^3$ corresponding to Definition~\ref{grp}:

\begin{definition}\label{oplus}
Given ${\cal P} \in \mathbb{Z}^3$ and ${\cal Q} \in \mathbb{Z}^3$, 
${\cal P} \oplus {\cal Q} \in \mathbb{Z}^3$ is defined in two cases:

\begin{itemize}
\item[(1)] If ${\cal P} = {\cal Q} = (m,n,z)$, then ${\cal P} \oplus {\cal Q} = (m', n', z')$,
where
\begin{eqnarray*}
z' & = & z_{dbl}({\cal P}) = 2nz,\\
w' & = & w_{dbl}({\cal P}) = 3m^2 + 2Amz^2 + z^4,\\
m' & = & m_{dbl}({\cal P}) = w'^2 - 4n^2(Az^2 + 2m),\\
n' & = & n_{dbl}({\cal P}) = w'(4mn^2 - m') - 8n^4.
\end{eqnarray*}

\item[(2)] If ${\cal P} = (x, y, 1) \in \mathbb{Z}^3$ and ${\cal Q} = (m,n,z) \neq {\cal P}$,
then ${\cal P} \oplus {\cal Q} = (m', n', z')$, where
\begin{eqnarray*}
z' & = & z_{sum}({\cal P},{\cal Q}) = z(z^2x - m),\\
m' & = & m_{sum}({\cal P},{\cal Q}) = \left(z^3y - n\right)^2 - \left(z^2(A+x) + m\right) \left(z^2x - m\right)^2\\
n' & = & n_{sum}({\cal P},{\cal Q}) = \left(z^3y - n\right)\left(z'^2x - m'\right) - z'^3y.
\end{eqnarray*}

\end{itemize}
\end{definition}

\begin{lemma}\label{decodecirc}
Let $P = \mathit{decode}({\cal P}) \in EC$ and $Q  = \mathit{decode}({\cal Q}) \in EC$, where ${\cal P} \oplus {\cal Q}$
is defined and if $P = Q$, then $P \neq O$ and ${\cal P} = {\cal Q}$.  Then $\mathit{decode}({\cal P} \oplus {\cal Q}) = P \oplus Q$.
\end{lemma}

{\sc Proof}:  The arithmetic operations below are to be understood as operations in $\mathbb{F}_\wp$ on the
field elements corresponding to the integers involved.

We first consider the case ${\cal P} = {\cal Q} = (m, n, z)$.  Let
\[
\lambda = \frac{3\left(\frac{m}{z^2}\right)^2 + 2A\left(\frac{m}{z^2}\right) + 1}{2\left(\frac{n}{z^3}\right)}
 = \frac{3m^2 + 2Amz^2 + z^4}{2nz} = \frac{w'}{z'}.
\]
Then $P \oplus P = (x, y)$, where
\[
x = \lambda^2 - A - 2\left(\frac{m}{z^2}\right) 
= \frac{w'^2}{z'^2} - \frac{Az^2 + 2m}{z^2} 
= \frac{w'^2}{z'^2} - \frac{4n^2(Az^2 + 2m)}{z'^2}
= \frac{m'}{z'^2}
\]
and
\[
y = \lambda\left(\frac{m}{z^2} - x\right) - \frac{n}{z^3} 
= \frac{w'}{z'}\cdot\frac{4mn^2 - m'}{z'^2} - \frac{8n^4}{z'^3} 
= \frac{w'(4mn^2 - m') - 8n^4}{z'^3}
= \frac{n'}{z'^3}.
\]
Thus, $\mathit{decode}({\cal P} \oplus {\cal P}) = \mathit{decode}(m',n',z') = (x,y) = P \oplus P$.

In the remaining case, we have ${\cal P} = (m, n, z)$ and ${\cal Q} = (x, y, 1)$.  Let
\[
\lambda = \frac{y - \frac{n}{z^3}} {x - \frac{m}{z^2}} = \frac{z^3y - n}{z(z^2x - m)} = \frac{z^3y - n}{z'},
\]
Then $P \oplus Q = (x', y')$, where
\begin{eqnarray*}
x' & = & \lambda^2 - A - x - \left(\frac{m}{z^2}\right) 
= \frac{(z^3y - x)^2}{z'^2} - \frac{z^2(A+x) + m}{z^2}\\
& = & \frac{(z^3y - x)^2}{z'^2} - \frac{\left(z^2(A+x) + m\right)\left(z^2x - m\right)^2}{z'^2}
= \frac{m'}{z'^2}
\end{eqnarray*}
and
\[
y' = \lambda(x - x') - y
= \frac{z^3y - n}{z'} \cdot \frac{xz'^2 - m'}{z'^2} - \frac{z'^3y}{z'^3}
= \frac{n'}{z'^3}.
\]
Thus, $\mathit{decode}({\cal P} \oplus {\cal Q}) = \mathit{decode}(m',n',z') = (x',y') = P \oplus Q$.~$\Box$\medskip

\section{Polynomial Terms and Sparse Horner Normal Form}\label{horner}

In this section, we describe our formalization of sparse Horner forms as S-expressions.
For this purpose, an S-expression is an integer, a symbol, or an ordered list $s = \mbox{\tt(}s_0\;s_1\;\ldots\;s_n\mbox{\tt )}$
of S-expressions.  In the last case, $\mathit{head}(s) = s_0$, and for $k \in \mathbb{N}$, we define
$s^{(k)} = \mbox{\tt(}s_k\;\ldots\;s_n\mbox{\tt )}$.  The set of all lists whose members are confined to a set $S$ is
${\cal L}(S)$.

Under the usual ACL2 encoding of multi-variable polynomials, a {\it polynomial term} over a list $V$ of variable symbols
is an S-expression constructed from integers and symbols in $V$ using the
symbols {\tt +}, {\tt -}, {\tt *}, and {\tt EXPT}.  The function {\it evalp} evaluates a term according
to an alist that associates variables with integer values in the natural way.  For example, if $V = \mbox{\tt (X Y Z)}$
and $A = \mbox{\tt ((X}\;\;2\mbox{\tt ) (Y}\;\;3\mbox{\tt ) (Z}\;\;0\mbox{\tt ))}$, then
$
\tau = \mbox{\tt (* X (EXPT (+ Y Z) 3))}
$
is a term over $V$ and $\mathit{evalp}(\tau,A) = 2\cdot (3+0)^3 = 54$.  The set of all polynomial terms over $V$ 
is denoted ${\cal T}(V)$.

We shall represent polynomial terms as objects of the following type:

\begin{definition}
A {\it sparse Horner form} (SHF) is any of the following:
\begin{itemize}
\item[(a)] An integer;
\item[(b)] A list \verb!(POP !$i\;\;p$\verb!)!, where $i \in \mathbb{N}$ and $p$ is a SHF
\item[(c)] A list \verb!(POW !$i\;\;p\;\;q$\verb!)!, where $i \in \mathbb{N}$ and $p$ and $q$ are SHFs.
\end{itemize}
A SHF is normal if its components are normal and it is not either of the following:
\begin{itemize}
\item[(a)] \verb!(POP !$i\;\;p$\verb!)!, where $i=0$ or $p \in \mathbb{Z}$ or $p = $\verb!(POP !$j\;\;q$\verb!)!;
\item[(b)] \verb!(POW !$i\;\;p\;\;q$\verb!)!, where $i=0$ or $p =$\verb!(POW !$j\;\;r\;\;0$\verb!)!.
\end{itemize}
${\cal H}$ denotes the set of all normal SHFs, or SHNFs.
\end{definition}

The evaluation of a SHF with respect to a list of integers is defined as follows:

\begin{definition}\label{evalh}
Let h be a SHF and let $N \in {\cal L}(\mathit{Z})$.
\begin{itemize}
\item[(a)] If $h \in \mathbb{Z}$, then $\mathit{evalh}(h,N) = h$.
\item[(b)] If $h = \mbox{\tt (POP }i\;\;p\mbox{\tt )}$, then then $\mathit{evalh}(h,N) = \mathit{evalh}(p,N^{(i)})$.
\item[(c)] If $h = \mbox{\tt (POW }i\;\;p\;\;q\mbox{\tt )}$ and $\mathit{head}(N) = n$, then
$\mathit{evalh}(h,N) = n^i\mathit{evalh}(p,N) + \mathit{evalh}(q,N^{(1)})$.
\item[(d)] If $h = \mbox{\tt (POW }i\;\;p\;\;q\mbox{\tt )}$ and $N = ${\tt ()}, then $\mathit{evalh}(h,N) = 0$.
\end{itemize}
\end{definition}

Our objective is to define, for a given variable list $V = $\verb!(!$v_0 \ldots v_k$\verb!)!, a mapping
$\mathit{norm} : {\cal T}(V) \rightarrow {\cal H}$
such that if $N =$ \verb!(!$n_0\ldots n_k$\verb!)! and $A = $\verb!((!$v_0\;\;n_0$\verb!)!$\ldots$\verb!(!$v_k\;\;n_k$\verb!))!,
then
\[
\mathit{evalh}(\mathit{norm}(x,V),N) = \mathit{evalp}(x,A).
\]
Thus, if two polynomials produce the same normal form, then they are equivalent.

A possible top-down approach to the definition of $\mathit{norm}(f, V)$ is as follows:

\begin{itemize}
\item[(1)] If $f$ is an integer constant, then $\mathit{norm}(f,V) = f$.
\item[(2)] Suppose $v_0$ occurs in $f$.  Find polynomials $g$ and $h$ such that $f = v_0^i\cdot g + h$, $g$ is not divisible by $v_0$,
and $v_0$ does not occur in $h$.  If $p=\mathit{norm}(g,V)$ and $q=\mathit{norm}(h,V^{(1)})$, then
\[
\mathit{norm}(f,V) = \mbox{\tt (POW }i\;\;p\;\;q\mbox{\tt )}.
\]
\item[(3)] Suppose $v_0$ does not occur in $f$.  Let $v_i$ be the first variable in $V$ that does occur in $f$.  If
  $p = \mathit{norm}(f,V^{(i)})$, then
\[
\mathit{norm}(f,V) = \mbox{\tt (POP }i\;\;p\mbox{\tt )}.
\]
\end{itemize}
For example, consider the polynomial $4x^4y^2 + 3x^3 + 2z^4 + 5$ with variable ordering $\mbox{\tt (}x\;\;y\;\;z\mbox{\tt )}$.
Rewriting the polynomial as
\[
x^3(4xy^2 + 3) + (2z^4 + 5),
\]
we find that the normalization is $\mbox{\tt (POW }3\;\;p\;\;q\mbox{\tt )}$, where
$p = \mathit{norm}(4xy^2 + 3, \mbox{\tt (}x\;\;y\;\;z\mbox{\tt )})$
and $q = \mathit{norm}(2z^4 + 5, \mbox{\tt (}y\;\;z\mbox{\tt )})$.
Continuing recursively, we arrive at the final result:
\begin{verbatim}
                  (POW 3 (POW 1 (POP 1 (POW 2 4 0)) 3)
                         (POP 1 (POW 4 2 5))).
\end{verbatim}
It may be instructive to check that the value of this SHF for the list of values $(1\;\;2\;\;3)$, 
for example, and the value of the represented polynomial for the corresponding alist, are both 207. 

It is not difficult to see that a SHF generated by this procedure is indeed normal.
Unfortunately, this approach is impractical because of the general difficulty of constructing the
polynomials $g$ and $h$ in Case (2).  Our preferred definition will provide a more efficient bottom-up procedure.
We begin with the two basic normalizing functions {\it pop} and {\it pow}:

\begin{definition}
Let $i \in \mathbb{N}$ and $p \in {\cal H}$.
\begin{itemize}
\item[(a)] If $i=0$ or $p \in \mathbb{Z}$, then $\mathit{pop}(i, p) = p$.
\item[(b)] If $p = \mbox{\tt (POP }j\;\;q\mbox{\tt )}$, then $\mathit{pop}(i, p) = \mbox{\tt (POP }i+j\;\;q\mbox{\tt )}$.
\item[(c)] Otherwise, $\mathit{pop}(i, p) = \mbox{\tt (POP }i\;\;p\mbox{\tt )}$.
\end{itemize}
\end{definition}

\begin{definition}
Let $i \in \mathbb{N} - \{0\}$, $p \in {\cal H}$, and $q \in {\cal H}$.
\begin{itemize}
\item[(a)] If $p=0$, then $\mathit{pow}(i, p, q) = \mathit{pop}(1, q)$.
\item[(b)] If $p = \mbox{\tt (POW }j\;\;r\;\;0\mbox{\tt )}$, then $\mathit{pow}(i, p, q) = \mbox{\tt (POW }i+j\;\;r\;\;q\mbox{\tt )}$.
\item[(c)] Otherwise, $\mathit{pow}(i, p, q) = \mbox{\tt (POW }i\;\;p\;\;q\mbox{\tt )}$.
\end{itemize}
\end{definition}

The following properties of these functions are immediate consequences of the definitions:

\begin{lemma}\label{poppow}
Let $i \in \mathbb{N}$, $p \in {\cal H}$, $q \in {\cal H}$, and $N \in {\cal L}(\mathit{Z})$.
\begin{itemize}
\item[(a)] $\mathit{pop}(i, p) \in {\cal H}$ and
$\mathit{evalh}(\mathit{pop}(i, p), N) = \mathit{evalh}(\mbox{\tt (POP }i\;\;p\mbox{\tt )}, N)$.
\item[(b)] If $i\neq 0$, then $\mathit{pow}(i, p, q) \in {\cal H}$ and
$\mathit{evalh}(\mathit{pow}(i, p, q), N) = \mathit{evalh}(\mbox{\tt (POW }i\;\;p\;\;q\mbox{\tt )}, N)$.
\end{itemize}
\end{lemma}

We also define a ring structure on ${\cal H}$,
Once we have computed the SHNFs for polynomials $x$ and $y$, the ring operations ``$\oplus$'' and ``$\otimes$''
compute those for $\mbox{\tt (+  }x\;\;y\mbox{\tt )}$ and $\mbox{\tt (*  }x\;\;y\mbox{\tt )}$.

\begin{definition}Let $x \in {\cal H}$ and $x \in {\cal H}$.
\begin{itemize}

\item[(1)] If $x \in \mathbb{Z}$, then
\begin{itemize}
\item[(a)] $y  \in \mathbb{Z} \Rightarrow x \oplus y = x+y$ and $x \otimes y = xy$.
\item[(b)] $y  = \mbox{\tt (POP  }i\;\;p\mbox{\tt )} \Rightarrow x \oplus y = \mbox{\tt (POP  }i\;\;x \oplus p\mbox{\tt )}$
and $x \otimes y = \mathit{pop}(i, x \otimes p)$.
\item[(c)] $y  = \mbox{\tt (POW  }i\;\;p\;\;q\mbox{\tt )} \Rightarrow x \oplus y = \mbox{\tt (POW  }i\;\;p\;\;x \oplus q\mbox{\tt )}$ and $x \otimes y = \mathit{pow}(i, x \otimes p, x \otimes q)$.
\end{itemize}

\item[(2)] If $y \in \mathbb{Z}$, then $x \oplus y = y \oplus x$ and $x \otimes y = y \otimes x$.

\item[(3)] If $x = \mbox{\tt (POP  }i\;\;p\mbox{\tt )}$ and $y = \mbox{\tt (POP  }j\;\;q\mbox{\tt )}$, then
\begin{itemize}
\item[(a)] $i=j \Rightarrow x \oplus y = \mathit{pop}(i, p \oplus q)$ and $x \otimes y = \mathit{pop}(i, p \otimes q)$.
\item[(b)] $i>j \Rightarrow x \oplus y = \mathit{pop}(j, \mbox{\tt (POP  }i-j\;\;p\mbox{\tt )} \oplus q)$ and
$x \otimes y = \mathit{pop}(j, \mbox{\tt (POP  }i-j\;\;p\mbox{\tt )} \otimes q)$.
\item[(c)] $i<j \Rightarrow x \oplus y = \mathit{pop}(i, \mbox{\tt (POP  }j-i\;\;q\mbox{\tt )} \oplus p)$ and
$x \otimes y = \mathit{pop}(i, \mbox{\tt (POP  }j-i\;\;q\mbox{\tt )} \otimes p)$.
\end{itemize}

\item[(4)] If $x = \mbox{\tt (POP  }i\;\;p\mbox{\tt )}$ and $y = \mbox{\tt (POW  }j\;\;q\;\;r\mbox{\tt )}$, then
\begin{itemize}
\item[(a)] $i=1 \Rightarrow x \oplus y = \mbox{\tt (POW  }j\;\;q\;\;r \oplus p\mbox{\tt )}$ and
$x \otimes y = \mbox{\tt (POW  }j\;\;x \otimes q\;\;p \otimes r\mbox{\tt )}$.
\item[(b)] $i>1 \Rightarrow x \oplus y = \mbox{\tt (POW  }j\;\;q\;\;r \oplus \mbox{\tt (POP  }i-1\;\;p\mbox{\tt ))}$ and
$x \otimes y = \mbox{\tt (POW  }j\;\;x \otimes q\;\;\mbox{\tt (POP  }i-1\;\;p\mbox{\tt )}\otimes r{\tt )}$.
\end{itemize}

\item[(5)] If $y = \mbox{\tt (POP  }i\;\;p\mbox{\tt )}$ and $y = \mbox{\tt (POW  }j\;\;q\;\;r\mbox{\tt )}$, then 
$x \oplus y = y \oplus x$ and $x \otimes y = y \otimes x$.

\item[(6)] If $x = \mbox{\tt (POW  }i\;\;p\;\;q\mbox{\tt )}$ and $y = \mbox{\tt (POW  }j\;\;r\;\;s\mbox{\tt )}$, then
\begin{itemize}
\item[(a)] $i=j \Rightarrow x \oplus y = \mathit{pow}(i, p \oplus r, q \oplus s)$.
\item[(b)] $i>j \Rightarrow x \oplus y = \mathit{pow}(j, \mbox{\tt (POW  }i-j\;\;p\;\;0\mbox{\tt )} \oplus r, q \oplus s)$
\item[(c)] $i<j \Rightarrow x \oplus y = \mathit{pow}(i, \mbox{\tt (POW  }j-i\;\;q\;\;0\mbox{\tt )} \oplus p, s \oplus q)$.
\item[(d)] $x \otimes y = \left(\mathit{pow}(i+j, p \otimes r, q \otimes s) \oplus \mathit{pow}(i, p \otimes \mathit{pop}(1, s), 0)\right)
\oplus \mathit{pow}(i, r \otimes \mathit{pop}(1, q), 0)$.
\end{itemize}

\end{itemize}
\end{definition}

The definitions of negation and exponentiation are straightforward:

\begin{definition}Let $x \in {\cal H}$.
\begin{itemize}
\item[(1)] If $x \in \mathbb{Z}$, then $\ominus x = -x$.
\item[(2)] If $x = \mbox{\tt (POP  }i\;\;p\mbox{\tt )}$, then $\ominus x = \mbox{\tt (POP  }i\;\;\ominus p\mbox{\tt )}$.
\item[(3)] If $x = \mbox{\tt (POW  }i\;\;p\;\;q\mbox{\tt )}$, then $\ominus x = \mbox{\tt (POW  }i\;\;\ominus p\;\;\ominus q\mbox{\tt )}$.
\end{itemize}
\end{definition}

\begin{definition}
If $x \in {\cal H}$ and $k \in \mathbb{N}$, then 
\[
x^k = \left\{\begin{array}{ll}
  1 & \mbox{if $k=0$}\\
  x \otimes x^{k-1} & \mbox{if $k>0$.}\end{array}\right.
\]
\end{definition}

The following properties are easily proved by induction:

\begin{lemma}\label{addl}
Let $x \in {\cal H}$, $y \in {\cal H}$, $N \in {\cal L}(\mathit{Z})$, and $k \in \mathbb{N}$.
\begin{itemize}
\item[(a)] $x \oplus y \in {\cal H}$ and
$\mathit{evalh}(x \oplus y, N) = \mathit{evalh}(x, N) + \mathit{evalh}(y, N)$.
\item[(b)] $x \otimes y \in {\cal H}$ and
$\mathit{evalh}(x \otimes y, N) = \mathit{evalh}(x, N) \cdot \mathit{evalh}(y, N)$.
\item[(c)] $\ominus x \in {\cal H}$ and $\mathit{evalh}(\ominus x, N) = -\mathit{evalh}(x, N)$.
\item[(d)] $x^k \in {\cal H}$ and $\mathit{evalh}(x^k, N) = \mathit{evalh}(x, N)^k$.
\end{itemize}
\end{lemma}

We can now define the normalization procedure:

\begin{definition}\label{norm}
If $x \in {\cal T}(V)$, where $V=\mbox{\tt (}v_0\ldots v_{k-1}\mbox{\tt )}$ is a list of distinct symbols, then

\begin{itemize}
\item[(1)] $x \in \mathbb{Z} \Rightarrow \mathit{norm}(x,V) = x$.
\item[(2)] $x =v_i$, $0 \leq i < k \Rightarrow \mathit{norm}(x,V) = \mathit{pop}(i, \mbox{\tt (POW  }1\;\;1\;\;0\mbox{\tt )})$.
\item[(3)] $x = \mbox{\tt (-  }y\mbox{\tt )} \Rightarrow \mathit{norm}(x,V) = \ominus \mathit{norm}(y,V)$.
\item[(4)] $x = \mbox{\tt (+  }y\;\;z\mbox{\tt )} \Rightarrow \mathit{norm}(x,V) = \mathit{norm}(y,V) \oplus \mathit{norm}(z,V)$.
\item[(5)] $x = \mbox{\tt (-  }y\;\;z\mbox{\tt )} \Rightarrow \mathit{norm}(x,V) = \mathit{norm}(y,V) \oplus (\ominus \mathit{norm}(z,V))$.
\item[(6)] $x = \mbox{\tt (*  }y\;\;z\mbox{\tt )} \Rightarrow \mathit{norm}(x,V) = \mathit{norm}(y,V) \otimes \mathit{norm}(z,V)$.
\item[(7)] $x = \mbox{\tt (EXPT  }y\;\;k\mbox{\tt )} \Rightarrow \mathit{norm}(x,V) = \mathit{norm}(y,V)^k$.
\end{itemize}
\end{definition}

The reader may wish to check that the SHNF for the polynomial $-z + x^3(z+x-3y)$ with respect to the variable list 
$\mbox{\tt (}x\;\;y\;\;z\mbox{\tt )}$ is once again
\begin{verbatim}
                  (POW 3 (POW 1 (POP 1 (POW 2 4 0)) 3)
                         (POP 1 (POW 4 2 5))).
\end{verbatim}

\begin{lemma}\label{evalhnorm}
Let $f \in {\cal T}(V)$, where $V=\mbox{\tt (}v_0\ldots v_{k-1}\mbox{\tt )}$ is a list of distinct symbols.
Let $N = \mbox{\tt (}n_0\ldots n_{\ell-1}\mbox{\tt )}$ be a list of integers with $\ell \geq k$ and
\[
A = \mbox{\tt ((}v_0\;\;n_0\mbox{\tt )}\ldots\mbox{\tt (}v_{k-1}\;\;n_{k-1}\mbox{\tt ))},
\]
Then $\mathit{norm}(f, V) \in {\cal H}$ and
\[
\mathit{evalh}(\mathit{norm}(f, V), N) = \mathit{evalp}(f, A).
\]
\end{lemma}

{\sc Proof}: The case $f = v_i$ follows from Definition~\ref{evalh} and Lemma~\ref{poppow}; the other cases follow from
Definition~\ref{evalh}, induction, and Lemma~\ref{addl}.~$\Box$\medskip

\section{Polynomial Reduction}\label{reduction}

We shall focus on the case of a list of variables corresponding to the coordinates of the points $P_0$, $P_1$, and $P_2$,
as characterized in Section~\ref{curve}.  We define the following lists:

\begin{definition}
${\cal V} = $\verb!(Y0 Y1 Y2 X0 X1 X2)!, ${\cal N} = $\verb!(!$y_0\;\;y_1\;\;y_2\;\;x_0\;\;x_1\;\;x_2$\verb!)!, and
\begin{center}
${\cal A} = $\verb!((Y0 !$y_0$\verb!) (Y1! $y_1$\verb!) (Y2 !$y_2$\verb!) (X0 !$x_0$\verb!) (X1 !$x_1$\verb!) (X2 !$x_2$\verb!))!
\end{center}
We abbreviate ${\cal T}({\cal V})$ as ${\cal T}$, and for $\tau \in {\cal T}$ we abbreviate 
$\mathit{evalp}(\tau, {\cal A})$ as $\hat{\tau}$.
\end{definition}

The ordering of the variable list ${\cal V}$ is designed to maximize the efficiency of the rewriting procedure
defined below.  This procedure operates on a SHF that represents a polynomial with respect to ${\cal V}$, which is 
effectively reduced, using the curve equation as a rewrite rule, to a polynomial that (a)~has the same value (modulo $\wp$)
as the given polynomial under the variable assignments of ${\cal A}$ and (b)~is at most linear in each of the variables {\tt Y}$i$.

The core of the rewriter is the function {\it split}, which reduces and splits a polynomial term $\tau$ into a sum of two polynomials, 
of which one is independent of a given {\tt Y}$j$ and the other is linear in {\tt Y}$j$.  More precisely, if 
$h = \mathit{norm}(\tau, {\cal V}^{(k)})$ and $0 \leq k \leq j \leq 3$, then $\mathit{split}(h,j,k)$ computes a pair of SHNFs
that represent these polynomials.

The following SHNF is used in the reduction:

\begin{definition}
$\Theta = $\verb!(POP !$3$\verb! (POW !$1$\verb! (POW !$1$\verb! (POW !$1\;\;1\;\;A$\verb!) !$1$\verb!) !$0$\verb!))!.
\end{definition}

\begin{lemma}
If $j \in \{0,1,2\}$, then $\mathit{evalh}(\Theta,{\cal N}^{(j)}) = x_j^3  + Ax_j^2 + x_j \equiv y_j^2 \pmod \wp$.
\end{lemma}

{\sc Proof}: This may be derived by expanding the definition of {\it evalh}.~$\Box$

\begin{definition}
Let $h \in {\cal H}$, $j \in \{0,1,2\}$, and $k \in \mathbb{N}$.
\begin{itemize}

\item[(1)] If $h \in \mathbb{Z}$ or $j<k$, then $\mathit{split}(h,j,k) = (h,0)$.

\item[(2)] If $j \geq k$, $h = $\verb!(POP !$i\;\;p$\verb!)!, and $(p_0,p_1) = \mathit{split}(p,j,k+i)$, then
\[
\mathit{split}(h,j,k) = (\mathit{pop}(i,p_0),\mathit{pop}(i,p_1)).
\]

\item[(3)] Let $h = $\verb!(POW!$\;i\;p\;q$\verb!)!, $(p_0,p_1) = \mathit{split}(p,j,k)$, and $(q_0,q_1) = \mathit{split}(q,j,k+1)$.

\begin{itemize}

\item[(a)] If $j>k$, then
\[
\mathit{split}(h,j,k) = \left(\mathit{pow}(i,p_0,q_0),\mathit{pow}(i,p_1,q_1)\right);
\]

\item[(b)] If $j=k$ and $i$ is even, then
\[
\mathit{split}(h,j,k) = 
\left((\Theta^{\frac{i}{2}} \otimes p_0) \oplus \mathit{pop}(1,q_0),
(\Theta^{\frac{i}{2}} \otimes p_1) \oplus \mathit{pop}(1,q_1)\right);
\]

\item[(c)] If $j=k$ and $i$ is odd, then
\[
\mathit{split}(h,j,k) = 
\left((\Theta^{\frac{i+1}{2}} \otimes p_1) \oplus \mathit{pop}(1,q_0),
(\Theta^{\frac{i-1}{2}} \otimes p_0) \oplus \mathit{pop}(1,q_1)\right).
\]

\end{itemize}

\end{itemize}
\end{definition}

\begin{lemma}\label{splitlemma}
Let $(h_0,h_1) = \mathit{split} (h, j, k)$, where $h \in {\cal H}$, $j \in \{0,1,2\}$, and $k \in \mathbb{N}$.  Then
$\mathit{evalh}(h, {\cal N}^{(k)}) 
\equiv \mathit{evalh}(h_0, {\cal N}^{(k)}) + y_j \cdot \mathit{evalh}(h_1, {\cal N}^{(k)}) \pmod \wp$.
\end{lemma}

{\sc Proof}: We may assume that $j \geq k$; otherwise
the claim is trivial.  The proof is by induction on the structure of $h$.
The case $h = $\verb!(POP !$i\;\;p$\verb!)! is straightforward:
\begin{eqnarray*}
\mathit{evalh}(h, {\cal N}^{(k)}) & = & \mathit{evalh}(p, {\cal N}^{(k+i)})\\
& \equiv & \mathit{evalh}(p_0, {\cal N}^{(k+i)}) + y_j \cdot \mathit{evalh}(p_1, {\cal N}^{(k+i)})\\
& = & \mathit{evalh}(\mathit{pop}(i, p_0), {\cal N}^{(k)}) + y_j \cdot \mathit{evalh}(\mathit{pop}(i, p_1), {\cal N}^{(k)})\\
& = & \mathit{evalh}(h_0, {\cal N}^{(k)}) + y_j \cdot \mathit{evalh}(h_1, {\cal N}^{(k)}).
\end{eqnarray*}

Suppose $h = $\verb!(POW !$i\;\;p\;\;q$\verb!)!.  By the definition of {\it evalh},
\begin{eqnarray*}
\lefteqn{\mathit{evalh}(h, {\cal N}^{(k)})}\\
& = & y_k^i \cdot \mathit{evalh}(p, {\cal N}^{(k)}) + \mathit{evalh}(q, {\cal N}^{(k+1)})\\
& \equiv & y_k^i \left(\mathit{evalh}(p_0, {\cal N}^{(k)}) + y_j \cdot \mathit{evalh}(p_1, {\cal N}^{(k)})\right)
+ \left(\mathit{evalh}(q_0, {\cal N}^{(k+1)})  + y_j \cdot \mathit{evalh}(q_1, {\cal N}^{(k+1)})\right)\\
& = & \left(y_k^i \mathit{evalh}(p_0, {\cal N}^{(k)}) + \mathit{evalh}(q_0, {\cal N}^{(k+1)})\right)
+ y_j \cdot \left(y_k^i \mathit{evalh}(p_1, {\cal N}^{(k)}) + \mathit{evalh}(q_1, {\cal N}^{(k+1)})\right).
\end{eqnarray*}
If $j > k$, then this may be written as
\begin{eqnarray*}
\lefteqn{\mathit{evalh}(\mbox{\tt (POW }i\;\;p_0\;\;q_0\mbox{\tt )},{\cal N}^{(k)})
+  y_j \cdot \mathit{evalh}(\mbox{\tt (POW }i\;\;p_1\;\;q_1\mbox{\tt )},{\cal N}^{(k)})}\\
& = & \mathit{evalh}(\mathit{pow}(i,p_0,q_0),{\cal N}^{(k)})
+  y_j \cdot \mathit{evalh}(\mathit{pow}(i,p_1,q_1),{\cal N}^{(k)})\\
& = & \mathit{evalh}(h_0,{\cal N}^{(k)})
+  y_j \cdot \mathit{evalh}(h_1,{\cal N}^{(k)}).
\end{eqnarray*}
We may assume, therefore, that $j=k$.  If $i$ is even, then
\begin{eqnarray*}
\lefteqn{y_k^i \mathit{evalh}(p_0, {\cal N}^{(k)}) + \mathit{evalh}(q_0, {\cal N}^{(k+1)})}\\
& = & (y_k^2)^\frac{i}{2}\mathit{evalh}(p_0, {\cal N}^{(k)}) 
+ \mathit{evalh}(q_0, {\cal N}^{(k+1)})\\
& \equiv & \mathit{evalh}(\Theta, {\cal N}^{(k)})^\frac{i}{2} \mathit{evalh}(p_0, {\cal N}^{(k)}) 
+ \mathit{evalh}(\mbox{\tt (POP }1\;\;q_0\mbox{\tt )}, {\cal N}^{(k)})\\
& = & \mathit{evalh}(h_0,{\cal N}^{(k)}),
\end{eqnarray*}
and similarly, 
\[
y_j\left(y_k^i \mathit{evalh}(p_1, {\cal N}^{(k)}) + \mathit{evalh}(q_1, {\cal N}^{(k+1)})\right) 
= y_j \cdot \mathit{evalh}(h_1,{\cal N}^{(k)}).
\]
if $i$ is odd, then we may rearrange the above expression for $\mathit{evalh}(h, {\cal N}^{(k)})$ as
\begin{eqnarray*}
& & \left((y_j^2)^\frac{i+1}{2}\mathit{evalh}(p_1, {\cal N}^{(k)}) + \mathit{evalh}(q_0, {\cal N}^{(k+1)})\right)\\
& + & y_j \left((y_j^2)^\frac{i-1}{2}\mathit{evalh}(p_0, {\cal N}^{(k)}) + \mathit{evalh}(q_1, {\cal N}^{(k+1)})\right),
\end{eqnarray*}
which similarly reduces to $\mathit{evalh}(h_0, {\cal N}^{(k)}) + y_j \cdot \mathit{evalh}(h_1, {\cal N}^{(k)})$.~$\Box$

\begin{definition}
If $h \in {\cal H}$, $j \in \{0,1,2\}$, and $(h_0,h_1) = \mathit{split}(h,j,0)$, then
\[
\mathit{rewrite}(h,j) = h_0 \oplus (h_1 \otimes \mathit{norm}(\mbox{\tt Y}j, {\cal V})).
\]
\end{definition}

\begin{lemma}\label{rlemma1}
If $h \in {\cal H}$, $j \in \{0,1,2\}$, and $r = \mathit{rewrite}(h,j)$,
then $\hat{r} \equiv \hat{h} \pmod \wp$.
\end{lemma}

{\sc Proof}: We instantiate Lemma~\ref{splitlemma} with $k = 0$ and invoke Lemma~\ref{addl}.~$\Box$

\begin{definition}\label{reduce}
If $\sigma \in {\cal T}$, then
\[
\mathit{reduce}(\sigma) 
= \mathit{rewrite}(\mathit{rewrite}(\mathit{rewrite}(\mathit{norm}(\sigma,{\cal V}), 0), 1), 2).
\]
\end{definition}

\begin{lemma}\label{rlemma}
If $\mathit{reduce}(\sigma) = \mathit{reduce}(\tau)$, then
$\hat{\sigma} \equiv \hat{\tau} \pmod \wp$.
\end{lemma}

{\sc Proof}: This is a consequence of Lemmas~\ref{rlemma1} and~\ref{evalhnorm}).~$\Box$

\section{Encoding Points on the Curve as Term Triples}\label{terms}

The evaluation of terms induces a mapping from ${\cal T}^3$ to $\mathbb{F}_\wp^2$:

\begin{definition}
For $\Pi = (\mu, \nu, \zeta) \in {\cal T}^3$, $\hat{\Pi} = (\hat{\mu}, \hat{\nu}, \hat{\zeta})$
and if $\hat{\zeta}$ is not divisible by $\wp$, then $\mathit{decode}(\Pi) = \mathit{decode}(\hat{\Pi})$.
\end{definition}

Clearly, under the following definitions, $\mathit{decode}(\Omega) = O$ and $\mathit{decode}(\Pi_i) = P_i$.

\begin{definition}\label{pii}
$\Omega = (0,0,1)$ and for $i \in \{0,1,2\}$, $\Pi_i = (\mbox{\tt X}i, \mbox{\tt Y}i, 1)$.\\
\end{definition}

Definition~\ref{oplus} suggests a partial addition on ${\cal T}^3$ corresponding to the group operation on $EC$.
This in combination with normalization (Definition~\ref{norm}) and reduction (Definition~\ref{reduce})
will provide a practical means of establishing equivalence between expressions constructed from the above
points by nested applications of $\oplus$, while avoiding the intractable task of explicitly computing those expressions.

\begin{definition}\label{bullet}
For $\Pi \in {\cal T}^3$ and $\Lambda \in {\cal T}^3$, 
$\Pi \oplus \Lambda \in {\cal T}^3$ is defined in the following cases:
\end{definition}

\begin{itemize}
\item[(1)] {\it If} $\Pi = \Lambda = (\mu,\nu,\zeta)$, {\it then} $\Pi \oplus \Lambda = (\mu', \nu', \zeta')$, {\it where}

\begin{tabbing}
$\mbox{~~~~~~~}\!\zeta'$\= $= \zeta_{dbl}(\Pi)=$ \verb!(* !$2$\verb! (* !$\nu\;\;\zeta$\verb!))!,
\end{tabbing}

\begin{tabbing}
$\mbox{~~~~~~~}\!\omega = \omega_{dbl}(\Pi)=$ \verb!(+ !\= \verb!(* !$3$\verb! (EXPT !$\mu\;\;2$\verb!))!\\
\> \verb!(+ !\= \verb!(* !$2$\verb! (* !$A$\verb! (* !$\mu$\verb! (EXPT !$\zeta\;\;2$\verb!))))!\\
\>\> \verb!(EXPT !$\zeta\;\;4$\verb!)))!,
\end{tabbing}

\begin{tabbing}
$\mbox{~~~~~~~}\!\mu'$\= $ = \mu_{dbl}(\Pi)$\\
\>$=$ \verb!(- !\= \verb!(EXPT !$\omega'\;\;2$\verb!)!\\
\>\> \verb!(* !$4$\verb! (* (EXPT !$\nu\;\;2$\verb!) (+ (* !$A$\verb! (EXPT !$\zeta\;\;2$\verb!)) (* !$2\;\;\mu$\verb!))))!, 
\end{tabbing}

\begin{tabbing}
$\mbox{~~~~~~~}\!\nu' = \nu_{dbl}(\Pi)=$ \verb!(- !\=\verb!(* !$\omega'$\verb! (- (* !$4$\verb! (* (EXPT !$\nu\;\;2$\verb!))) !$\mu'$\verb!))!\\
\> \verb!(* !$8$\verb! (EXPT !$\nu\;\;4$\verb!))).!
\end{tabbing}

\item[(2)] {\it If} $\Pi = (\theta, \phi, 1)$ and $\Lambda = (\mu,\nu,\zeta) \neq \Pi$, 
{\it then} $\Pi \oplus \Lambda = (\mu', \nu', \zeta')$, {\it where}

\begin{tabbing}
$\mbox{~~~~~~~}\!\zeta' = \zeta_{sum}(\Pi, \Lambda)
=$ \verb!(* !$\zeta$\verb! (- (* (EXPT !$\zeta\;\;2$\verb!) !$\theta$\verb!) !$\mu$\verb!)!,
\end{tabbing}

\begin{tabbing}
$\mbox{~~~~~~~}\!\mu' = \mu_{sum}(\Pi, \Lambda)
=$ \verb!(- !\= \verb!(EXPT (- (* (EXPT !$\zeta\;\;3$\verb!) !$\nu$\verb!) !$2$\verb!)!\\
\> \verb!(* !\= \verb!(+ (* (EXPT !$\zeta\;\;2$\verb!) (+ !$A\;\;\theta$\verb!)) !$\mu$\verb!)!\\
\>\> \verb!(EXPT (- (* (EXPT !$\zeta\;\;2$\verb!) !$\theta$\verb!) !$\mu$\verb!) !$2$\verb!)))!,
\end{tabbing}

\begin{tabbing}
$\mbox{~~~~~~~}\!\nu'= \nu_{sum}(\Pi, \Lambda)
=$ \verb!(- !\=\verb!(* !\= \verb!(- (* (EXPT! $\zeta\;\;3$ \verb!) !$\phi$\verb!) !$\nu$\verb!)!\\
\> \verb!(- (* (EXPT !$\zeta'\;\;2$\verb!) !$\theta$\verb!) !$\mu'$\verb!))!\\
\>\> \verb!(* (EXPT !$\zeta\;\;3$\verb!) !$\phi$\verb!))!.
\end{tabbing}

\end{itemize}

\begin{lemma}\label{blemma}
Let $\Pi \in {\cal T}^3$ and $\Lambda \in {\cal T}^3$ with $\mathit{decode}(\Pi) = P \in EC$,
$\mathit{decode}(\Lambda) = Q \in EC$, and $\Pi \oplus \Lambda$ defined.  Assume that if
$\Pi = \Lambda$, then $P = Q \neq O$, and otherwise $P \neq Q$.  Then
$\mathit{decode}(\Pi \oplus \Lambda) = P \oplus Q$.
\end{lemma}

{\sc Proof}: Let $\Gamma = \Pi \oplus \Lambda$.  Clearly, the hypothesis implies that $\hat{\Pi} \oplus \hat{\Lambda}$
is defined.  In light of Lemma~\ref{decodecirc}, we need only show that  
$\hat{\Gamma} = \hat{\Pi} \oplus \hat{\Lambda}$.

We shall examine the case $\Pi = \Lambda$; the remaining case is similar.  Let $\Pi = (\mu, \nu, \zeta)$,
$\Lambda = (\mu', \nu', \zeta')$, and
\[
{\cal P} = \hat{\Pi} \oplus \hat{\Lambda} = (\hat{\mu}, \hat{\nu}, \hat{\zeta}) = (m, n, z).  
\]
According to Definition~\ref{bullet},
\begin{center}
$\zeta' = $ \verb! (* 2 (* ! $\nu \;\; \zeta$\verb!))!
\end{center}
and it is clear from the definition of {\it evalp} that
\[
 \hat{\zeta'} = \mathit{evalp}(\zeta', {\cal A})
= 2 \cdot \mathit{evalp}(\nu, {\cal A}) \cdot \mathit{evalp}(\zeta, {\cal A})
= 2\hat{\nu}\hat{\zeta}
= 2mn
= z_{dbl}({\cal P}).
\]
It may similarly be shown that $\hat{\mu} = m_{dbl}({\cal P})$ and $\hat{\nu} = n_{dbl}({\cal P})$.  Thus,
\[
\hat{\Gamma} = (\hat{\mu'}, \hat{\nu'}, \hat{\zeta'}) = (z_{dbl}({\cal P}), m_{dbl}({\cal P}), n_{dbl}({\cal P})) 
= {\cal P} \oplus {\cal P}.~\Box
\]
We also define a negation operator, with the obvious property:
\begin{definition}\label{neg}
For $\Pi = (\mu, \nu, \zeta) \in {\cal T}^3$, $\ominus\Pi = (\mu, \mbox{\tt (- }\nu\mbox{\tt )}, \zeta)$.
\end{definition}

\begin{lemma}\label{nlemma}
If $\Pi \in {\cal T}^3$ and $\mathit{decode}(\Pi) \in EC$, then $\mathit{decode}(\ominus\Pi) = \ominus P$.
\end{lemma}

The next two lemmas, which combine the results of this section with those of Section~\ref{reduction}, will be critical in 
establishing the group axioms: Lemma~\ref{elemma} for closure and Lemma~\ref{slemma} for commutativity and associativity.

\begin{definition}\label{ecenc}
Given $\Pi = (\mu, \nu, \zeta) \in {\cal T}^3$, let
\end{definition}
\begin{Verbatim}[commandchars=\|\[\]]
      |begin[math]|tau =|end[math] (- (EXPT |begin[math]|nu|end[math] |begin[math]2|end[math])
              (+ (EXPT |begin[math]|mu|end[math] |begin[math]3|end[math])
                 (+ (* |begin[math]A|end[math] (EXPT (* |begin[math]|mu|end[math] |begin[math]|zeta|end[math]) |begin[math]2|end[math]))
                    (* |begin[math]|mu|end[math] (EXPT |begin[math]|zeta|end[math] |begin[math]4|end[math]))))).
\end{Verbatim}
Then $\Pi$ {\it is an EC-encoding if} $\mathit{reduce}(\tau) = 0$.

\begin{lemma}\label{elemma}
If $\Pi$ is an EC-encoding and $P = \mathit{decode}(\Pi)$, then $P \in EC$.
\end{lemma}

{\sc Proof}:  Let $\Pi = (\mu, \nu, \zeta)$, $\hat{\Pi} = (m,n,z)$, and $P = (x,y) = \mathit{decode}(\Pi)$.  Then 
\[
\hat{\tau} = n^2 - (m^3 + A(mz)^2 + mz^4)
\]
and
\[
P = \left(\frac{\bar{m}}{\bar{z}^2}, \frac{\bar{n}}{\bar{z}^3}\right).
\]
By Lemma~\ref{rlemma}, $\hat{\tau} \equiv 0 \pmod \wp$, and therefore, in the field $\mathbb{F}_\wp$,
\[
\bar{n}^2 = \bar{m}^3 + A(\bar{m}\bar{z})^2 + \bar{m}\bar{z}^4.
\]
Dividing this equation by $\bar{z}^6$ yields
\[
y^2 = \frac{\bar{n}^2}{\bar{z}^6} 
= \frac{\bar{m}^3}{\bar{z}^6} + \frac{A\bar{m}^2}{\bar{z}^4} + \frac{\bar{m}}{\bar{z}^2} = x^3 + Ax^2 + x.~\Box
\]

\begin{definition}\label{sim}
Given $\Pi = (\mu, \nu, \zeta) \in {\cal T}^3$ and $\Pi' = (\mu', \nu', \zeta') \in {\cal T}^3$, let
\begin{center}
$\sigma =$\verb!(* !$\mu$\verb! (EXPT ! $\zeta'\;\;2$ \verb!))!,\mbox{~~~~}
$\sigma' =$\verb!(* !$\mu'$\verb! (EXPT ! $\zeta\;\;2$ \verb!))!,\\
$\tau =$\verb!(* !$\nu$\verb! (EXPT ! $\zeta'\;\;3$ \verb!))!,\mbox{~~~~}
$\tau' =$\verb!(* !$\nu$\verb! (EXPT ! $\zeta\;\;3$ \verb!))!.
\end{center}
Then $\Pi \sim \Pi' \Leftrightarrow \mathit{reduce}(\sigma) = \mathit{reduce}(\sigma') \mbox{ and }
\mathit{reduce}(\tau) = \mathit{reduce}(\tau')$.
\end{definition}

\begin{lemma}\label{slemma}
Let $\Pi \in {\cal T}^3$ and $\Pi' \in {\cal T}^3$.  If $\mathit{decode}(\Pi) = P \in EC$,
$\mathit{decode}(\Pi') = P' \in EC$, and $\Pi \sim \Pi'$, then $P = P'$.
\end{lemma}

{\sc Proof}: Let $\Pi = (\mu, \nu, \zeta)$, $\Pi' = (\mu', \nu', \zeta')$, $\hat{\Pi} = (m,n,z)$,
and $\hat{\Pi'} = (m',n',z')$.  Then by Lemma~\ref{rlemma},
\[
mz'^2 = \hat{\mu}\hat{\zeta'}^2 = \hat{\sigma} \equiv \hat{\sigma'} = \hat{\mu'}\hat{\zeta}^2 = m'z^2 \pmod \wp
\]
and
\[
nz'^3 = \hat{\nu}\hat{\zeta'}^3 = \hat{\tau} \equiv \hat{\tau'} = \hat{\nu'}\hat{\zeta}^3 = n'z^3 \pmod \wp.
\]
Thus, in the field $\mathbb{F}_\wp$,
\[
P = \left(\frac{\bar{m}}{\bar{z}^2}, \frac{\bar{n}}{\bar{z}^3}\right)
= \left(\frac{\bar{m'}}{\bar{z'}^2}, \frac{\bar{n'}}{\bar{z'}^3}\right) = P'.~\Box
\]

\section{Abelian Group Axioms}\label{axioms}

It must be shown that if $\{P, Q, R\} \subset EC$, then $P \oplus Q = Q \oplus P \in EC$
and $(P \oplus Q) \oplus R =  P \oplus (Q \oplus R)$.  We may assume that the points are
finite, since each of these properties is trivial otherwise, and without loss of generality,
we may confine our attention to the fixed points $P_0$, $P_1$, and $P_2$.  

Computations~\ref{p1}--\ref{p13} below are computational results of evaluating the functions
that are specified by Definitions~\ref{reduce} (term reduction), \ref{bullet} (addition of term triples),
\ref{neg} (negation of a term triple), \ref{ecenc} (EC-encoding recognizer), and~\ref{sim} (equivalence
of term triples).  The lemmas of this section are derived from these results
using the corresponding Lemmas~\ref{rlemma}, \ref{blemma}, \ref{nlemma}, \ref{elemma}, and~\ref{slemma}.

\begin{computation}\label{p1}
$\Pi_0 \oplus \Pi_0$ and $\Pi_0 \oplus \Pi_1$ are EC-encodings.
\end{computation}

\begin{lemma}[Closure]\label{closure}
$P_0 \oplus P_1 \in EC$.
\end{lemma}

{\sc Proof}: If $P_0 \neq P_1$, then by Computation~\ref{p1}, Definition~\ref{pii} and Lemmas~\ref{blemma},
\[
P_0 \oplus P_1 = \mathit{decode}(\Pi_0) \oplus \mathit{decode}(\Pi_1) = \mathit{decode}(\Pi_0 \oplus \Pi_1),
\]
and by Lemma~\ref{elemma}, $P_0 \oplus P_1 \in EC$.  Similarly, $P_0 \oplus P_0 \in EC$.~$\Box$

\begin{computation}\label{p2}
$\Pi_0 \oplus \Pi_1 \sim \Pi_1 \oplus \Pi_0$.
\end{computation}

\begin{lemma}[Commutativity]\label{commutativity}
$P_0 \oplus P_1 = P_1 \oplus P_0$.
\end{lemma}

{\sc Proof}: We may assume $P_0 \neq P_1$.  By Computation~\ref{p2} and Lemmas~\ref{blemma} and~\ref{slemma},
\[
P_0 \oplus P_1 = \mathit{decode}(\Pi_0 \oplus \Pi_1) = \mathit{decode}(\Pi_1 \oplus \Pi_0) = P_1 \oplus P_0.~\Box
\]

All remaining results pertain to associativity.

\begin{computation}\label{p3}
$\ominus(\Pi_0 \oplus \Pi_0) \sim (\ominus\Pi_0) \oplus (\ominus\Pi_0)$.
\end{computation}

\begin{computation}\label{p4}
$\ominus(\Pi_0 \oplus \Pi_1) \sim (\ominus\Pi_0) \oplus (\ominus\Pi_1)$.
\end{computation}

\begin{lemma}\label{c0}
$\ominus(P_0 \oplus P_1) =  (\ominus P_0) \oplus (\ominus P_1)$.
\end{lemma}

{\sc Proof}: This follows from Computations~\ref{p3} and~\ref{p4} and Lemmas~\ref{rlemma}, \ref{blemma}, \ref{nlemma}, 
and~\ref{slemma}.~$\Box$

\begin{computation}\label{p5}
$(\ominus\Pi_0) \oplus (\Pi_0 \oplus \Pi_0) \sim \Pi_0$.
\end{computation}

\begin{computation}\label{p6}
$(\ominus\Pi_0) \oplus (\Pi_0 \oplus \Pi_1) \sim \Pi_1.$
\end{computation}

\begin{lemma}\label{c2}
If $P_0 \oplus P_1 \neq \ominus P_0$, then $(\ominus P_0) \oplus (P_0 \oplus P_1) = P_1$.
\end{lemma}

{\sc Proof}: This follows similarly from Computations~\ref{p5} and~\ref{p6}.~$\Box$

\begin{computation}\label{p9}
$\Pi_2 \oplus (\Pi_0 \oplus \Pi_1) \sim \Pi_1 \oplus (\Pi_0 \oplus \Pi_2)$.
\end{computation}

\begin{computation}\label{p10}
$\Pi_1 \oplus (\Pi_0 \oplus \Pi_0) \sim \Pi_0 \oplus (\Pi_0 \oplus \Pi_1)$.
\end{computation}

\begin{lemma}\label{c4}
If $P_0 \oplus P_1 \notin \{P_2, \ominus P_2\}$ and $P_0 \oplus P_2 \notin \{P_1, \ominus P_1\}$, then
\[
P_2 \oplus (P_0 \oplus P_1) = P_1 \oplus (P_0 \oplus P_2).
\] 
\end{lemma}

{\sc Proof}: The claim follows immediately from Computations~\ref{p9} and~\ref{p10} and Lemmas~\ref{rlemma}, ~\ref{blemma}, 
and~\ref{slemma} except in the cases $P_0 = \ominus P_1$, $P_0 = P_1 = O$, $P_0 = \ominus P_2$, and $P_0 = P_2 = O$.  We need only consider 
the first two of these cases; the other two are similar.  Moreover, since $\ominus O = O$, the second case is subsumed by the first.  
Thus, we may assume $P_0 = \ominus P_1$. Now $LHS$ (the left-hand side) is $P_2$ and by Lemma~\ref{c2}, 
$RHS = (\ominus P_0) \oplus (P_0 \oplus P_2) = P_2 = LHS$.~$\Box$

\begin{computation}\label{p11}
$(\Pi_0 \oplus \Pi_0) \oplus (\Pi_0 \oplus \Pi_0) \sim \Pi_0 \oplus (\Pi_0 \oplus (\Pi_0 \oplus \Pi_0))$.
\end{computation}

\begin{computation}\label{p12}
$(\Pi_0 \oplus \Pi_1) \oplus (\Pi_0 \oplus \Pi_1) \sim \Pi_0 \oplus (\Pi_1 \oplus (\Pi_0 \oplus \Pi_1))$.
\end{computation}

\begin{lemma}\label{c5}
If $P_0 \oplus P_1 \neq -(P_0 \oplus P_1)$, $P_0 \oplus P_1 \neq \ominus P_1$, and 
$P_1 \oplus (P_0 \oplus P_1) \notin \{P_0,\ominus P_0\}$, then
$(P_0 \oplus P_1) \oplus (P_0 \oplus P_1) = P_0 \oplus (P_1 \oplus (P_0 \oplus P_1))$.
\end{lemma}

{\sc Proof}: The case $P_0 = \ominus P_1$ is trivial and the case $P_1 = P_0 \oplus P_1$ is precluded by Lemma~\ref{l1}.  All other cases are
handled by Computations~\ref{p11} and~\ref{p12} and Lemmas~\ref{rlemma}, \ref{blemma}, and~\ref{slemma}.~$\Box$

\begin{computation}\label{p13}
Let $\Sigma = \Pi_0 \oplus \Pi_1 = (\mu,\nu,\zeta)$, $\Sigma' = \Pi_0 \oplus \Pi_0 = (\mu',\nu',\zeta')$,
\begin{tabbing}
\mbox{~~~~~~~~}$\phi = $\verb!(- !\= \verb!(EXPT !\= \verb!(+ !\= \verb!(- !$\mu$\verb! (* X1 (EXPT !$\zeta\;\;2$\verb!))) (* !$2$\verb! (* Y1 Y2))) !$2$\verb!)!\\
\>\verb!(EXPT (* !$2$\verb! (* Y1 Y2)) !$2$\verb!))!,
\end{tabbing}
and
\begin{tabbing}
\mbox{~~~~~~~~}$\psi = $\verb!(* !\=\verb!(- !$\mu'$\verb! (* X2 (EXPT !$\zeta'\;\;2$\verb!))) (EXPT !$\zeta\;\;2$\verb!))!.
\end{tabbing}
Then $\mathit{reduce}(\phi) = \mathit{reduce}(\psi)$.
\end{computation}

\begin{lemma}\label{c6}
If $P_0 \oplus P_1 = \ominus P_0$, then $P_1 = \ominus(P_0 \oplus P_0)$.
\end{lemma}

{\sc Proof}: First note that we may assume that $P_0 \notin \{P_1,\ominus P_1\}$, for if $P_0 = P_1$, then
\[
P_1 = \ominus(\ominus P_0) = \ominus(P_0 \oplus P_1) = \ominus(P_0 \oplus P_0),
\]
and if $P_0 = \ominus P_1$, then
\[
\ominus P_0 = P_0 \oplus P_1 = (\ominus P_1) \oplus P_1 = \infty,
\]
contradicting $P_0 \neq \infty$.  Furthermore, if $P_0 = \ominus P_0$, then $P_0 \oplus P_1 = P_0$,
contradicting Lemma~\ref{l1}. Thus, we have $x_0 \neq x_1$ and $y_0 \neq 0$.

Retaining the notation of Computation~\ref{p13}, let
\[
\hat{\Sigma} = (\hat{\mu},\hat{\nu}, \hat{\zeta}) = (m,n,z)
\]
and
\[
\hat{\Sigma'} = (\hat{\mu'},\hat{\nu'}, \hat{\zeta'}) = (m',n',z').
\]
It follows from the definition of {\it evalp} that
\[
\hat{\phi} = (m-x_0z^2 + 2y_0y_1)^2 - (2y_0y_0)^2
\]
and
\[
\hat{\psi} = (m' - x_1z'^2)z^2.
\]
By Lemma~\ref{blemma},
\[
\mathit{decode}(\Sigma) = \left(\frac{\bar{m}}{\bar{z}^2}, \frac{\bar{n}}{\bar{z}^3}\right) = P_0 \oplus P_1 = \ominus P_0 = (x_0,-y_0),
\]
and hence $m \equiv x_0z^2 \pmod \wp$, which implies $\hat{\phi} \equiv 0 \pmod \wp$.

By Computation~\ref{p13} and Lemma~\ref{rlemma}, $\hat{\psi} \equiv 0 \pmod \wp$, which implies $m' \equiv x_1z'^2 \pmod \wp$.
Thus, by Lemma~\ref{blemma},
\[
P_0 \oplus P_0 = \mathit{decode}(\Sigma') = \left(\frac{\bar{m'}}{\bar{z'}^2}, \frac{\bar{n'}}{\bar{z'}^3}\right) 
= \left(x_1, \frac{\bar{n'}}{\bar{z'}^3}\right),
\]
which implies $P_0 \oplus P_0 \in \{P_1, \ominus P_1\}$.  We need only consider the case $P_0 \oplus P_0 = P_1$.

Suppose that $P_0 \oplus P_0 = P_1$.  Then $P_0 \oplus P_0 \neq \ominus P_0$, and by Lemma~\ref{c3},
\[
P_1 \oplus (\ominus P_0) = (P_0 \oplus P_0) \oplus (\ominus P_0) = P_0.
\]
Thus,
\[
P_0 \oplus (\ominus P_1) = \ominus(\ominus P_0 \oplus P_1) = \ominus(P_1 \oplus (\ominus P_0)) = \ominus P_0 = P_0 \oplus P_1.
\]
By Lemma~\ref{l2}, $P_1 = O$, and hence $P_1 = \ominus P_1 = \ominus(P_0 \oplus P_0)$.~$\Box$

\begin{lemma}\label{l3}
If $P_0 \oplus P_1 = \ominus P_2$, then $(P_0 \oplus P_1) \oplus P_2 = P_0 \oplus (P_1 \oplus P_2)$.
\end{lemma}

{\sc Proof}: $LHS = \infty$ and by Lemma~\ref{c0},
\[
 RHS = P_0 \oplus (P_1 \oplus (\ominus(P_0 \oplus P_1))) = P_0 \oplus (P_1 \oplus ((\ominus P_0) \oplus (\ominus P_1))).
\]
Therefore, we must show that $P_1 \oplus ((\ominus P_0) \oplus (\ominus P_1)) = \ominus P_0$.

If $(\ominus P_0) \oplus (\ominus P_1) \neq P_1$, then this follows from Lemma~\ref{c2}.
On the other hand, if $(\ominus P_0) \oplus (\ominus P_1) = P_1$, then by Lemmas~\ref{c6} and~\ref{c0},
\[
\ominus P_0 = \ominus((\ominus P_1) \oplus (\ominus P_1)) = P_1 \oplus P_1 = P_1 \oplus ((\ominus P_0) \oplus (\ominus P_1)).~\Box
\]

\begin{lemma}\label{l4}
$(P_0 \oplus P_0) \oplus P_1 = P_0 \oplus (P_0 \oplus P_1)$.
\end{lemma}

{\sc Proof}:  By Lemma~\ref{l3}, we may assume $P_0 \oplus P_1 \neq \ominus P_0$ and $P_0 \oplus P_0 \neq \ominus P_1$.
By Lemmas~\ref{l1} and~\ref{c4}, we may assume that $P_1 = P_0 \oplus P_1$.

If $P_1 = \ominus P_0$, then 
\[
LHS = P_1 + P_1 = (\ominus P_0) \oplus (\ominus P_0) = \ominus(P_0 \oplus P_0) = \ominus P_1 = P_0 = RHS.
\]
But if $P_1 \neq \ominus P_0$, then the claim follows from Lemma~\ref{c5}.~$\Box$\medskip

Two final computations are required for the case $P_1 = O$ and $P_2 = P_0 \oplus P_1$:

\begin{computation}\label{p8}
$(\Pi_0 \oplus \Omega) \oplus (\Pi_0 \oplus \Omega) \sim \Pi_0 \oplus \Pi_0$.
\end{computation}

\begin{computation}\label{p7}
$\Omega \oplus (\Pi_0 \oplus \Omega) \sim \Pi_0$.
\end{computation}

\begin{lemma}\label{c3}
$(P_0 \oplus O) \oplus (P_0 \oplus O) = P_0 \oplus (O \oplus (P_0 \oplus O))$.
\end{lemma}

{\sc Proof}: We may assume that $P_0 \neq O$.  Since Lemma~\ref{l1} implies $P_0 \oplus O \neq O$, it follows from Computation~\ref{p7}
that $O \oplus (P_0 \oplus O) = P_0$.  Thus, the claim reduces to $(P_0 \oplus O) \oplus (P_0 \oplus O) = P_0 \oplus P_0$,
which follows from Computation~\ref{p8}.~$\Box$

\begin{lemma}[Associativity]\label{associativity}
$(P_0 \oplus P_1) \oplus P_2 = P_0 \oplus (P_1 \oplus P_2)$.
\end{lemma}

{\sc Proof}: By Lemmas~\ref{c5} and~\ref{l3}, we may assume $P_1 \oplus P_2 \neq \ominus P_0$, and $P_2 = P_0 \oplus P_1$.
By Lemma~\ref{c5}, we need only eliminate the cases $P_0 \oplus P_1 = \ominus P_1$ and $P_1 \oplus (P_0 \oplus P_1) = P_0$.

If $P_2 = P_0 \oplus P_1 = \ominus P_1$, then $RHS = P_0$ and by Lemmas~\ref{c0} and~\ref{c6}, 
\[
LHS = (P_0 \oplus P_1) \oplus (P_0 \oplus P_1) = (\ominus P_1) \oplus (\ominus P_1) = \ominus(P_1 \oplus P_1) = P_0 = RHS.
\]
Finally, if $P_1 \oplus (P_0 \oplus P_1) = P_0$, then Lemma~\ref{l4} implies $P_0 = P_0 \oplus (P_1 \oplus P_1)$,
Lemma~\ref{l1} then implies $P_1 = O$, and the claim follows from Lemma~\ref{c3}.~$\Box$

\nocite{*}
\bibliographystyle{eptcs}
\bibliography{bib}
\end{document}